# Cosmology from Antarctica

Robert W. Wilson & Antony A. Stark

*Smithsonian Astrophysical Observatory, 60 Garden St., Cambridge, MA 02138*

**Abstract.** Four hundred thousand years after the Big Bang, electrons and nuclei combined to form atoms for the first time, allowing a sea of photons to stream freely through a newly-transparent Universe. After billions of years, those photons, highly redshifted by the universal cosmic expansion, have become the Cosmic Microwave Background Radiation (CMB) we see coming from all directions today. Observation of the CMB is central to observational cosmology, and the Antarctic Plateau is an exceptionally good site for this work. The first attempt at CMB observations from the Plateau was an expedition to the South Pole in December 1986 by the Radio Physics Research group at Bell Laboratories. No CMB anisotropies were observed, but sky noise and opacity were measured. The results were sufficiently encouraging that in the Austral summer of 1988-1989, three CMB groups participated in the "Cucumber" campaign, where a temporary site dedicated to CMB anisotropy measurements was set up 2 km from South Pole Station. These were summer-only campaigns. Winter-time observations became possible with the establishment in 1990 of the Center for Astrophysical Research in Antarctica (CARA), a National Science Foundation Science and Technology Center. CARA developed year-round observing facilities in the "Dark Sector", a section of Amundsen-Scott South Pole Station dedicated to astronomical observations. CARA scientists fielded several astronomical instruments: AST/RO, SPIREX, White Dish, Python, Viper, ACBAR, and DASI. By 2001, data from CARA, together with BOOMERANG, a CMB experiment on a long-duration balloon launched from McMurdo Station on the coast of Antarctica, showed clear evidence that the overall geometry of the Universe is flat, as opposed to being positively or negatively curved. In 2002, the DASI group reported the detection of polarization in the CMB. These observations strongly support a "Concordance Model" of cosmology, where the dynamics of a flat Universe are dominated by forces exerted by the mysterious Dark Energy and Dark Matter. CMB observations continue on the Antarctic Plateau. The South Pole Telescope (SPT) is a newly-operational 10 m diameter offset telescope designed to rapidly measure anisotropies on scales much smaller than $1°$.

## 1. Introduction

Cosmology has made tremendous strides in the past decade ─ this is generally understood within the scientific community, but it is not generally appreciated that some of the most important results have come from Antarctica. Observational cosmology has become a quantitative science. Cosmologists describe the Universe by a model with roughly a dozen parameters, for example the Hubble constant, $H_0$, and the density parameter, $\Omega$. A decade ago, typical errors on these parameters were 30% or greater; now, most are known within 10%. We can honestly discriminate for and against cosmological hypotheses on the basis of quantitative data. The current concordance model, Lambda-Cold Dark Matter (Ostriker and Steinhardt 1995), is both highly detailed and consistent with observations. This paper will review the contribution of Antarctic observations to this great work.

From the first detection of the Cosmic Microwave Background (CMB) radiation (Penzias & Wilson 1965), it was understood that deviations from perfect anisotropy would advance our understanding of cosmology (Peebles & Yu 1970, Harrison, 1970): the small deviations from smoothness in the early Universe are the seeds from which subsequent structure grows, and these small irregularities appear as differences in the brightness of the CMB in various directions on the sky. When the Universe was only 350,000 years old, the CMB radiation was released by electrons as they combined with nuclei into atoms for the first time. Anisotropies in the CMB radiation indicate slight differences in density and temperature that eventually evolve into stars, galaxies, and clusters of galaxies. Observations at progressively higher sensitivity by many groups of scientists from the 1970s through the 1990s failed to detect the anisotropy (cf. the review by Lasenby, Jones, & Drabowski 1998). In the course of these experiments, observing techniques were developed, detector sensitivities were improved by orders of magnitude, and the effects of atmospheric noise became better understood. The techniques and detectors were so improved that the sensitivity of experiments came to be dominated by atmospheric noise at most observatory sites. Researchers moved their instruments to orbit, to balloons, and to high, dry observatory sites in the Andes and in Antarctica. Eventually, CMB anisotropies were detected by the Cosmic Background Explorer satellite (COBE, Fixsen et al. 1996). The ground-based experiments at remote sites also met with success. The spectrum of brightness in CMB variations as a function of spatial frequency was measured by a series of ground-based and balloon-borne experiments, many of them located in the Antarctic. The data were then vastly improved upon by the Wilkinson Microwave Anisotropy Probe (WMAP, Spergel et al. 2003). The future Planck satellite mission (Tauber 2005), expected launch in 2008, will provide high signal-to-noise data on CMB anisotropy and polarization that will reduce the error on some cosmological parameters to the level of one percent.

Even in the era of CMB satellites, ground-based CMB observations are still essential for reasons of fundamental physics. CMB radiation occurs only at wavelengths longer than 1 millimeter. The resolution of a telescope (in radians) is equal to the observed wavelength divided by the telescope diameter. To work properly, the overall accuracy of the telescope optics must be a small fraction of a wavelength. Observing the CMB at resolutions of a minute of arc or smaller therefore requires a telescope that is 10 m in diameter or larger, with an overall accuracy of 0.1 millimeter or better. There are no prospects for an orbital or airborne telescope of this size and accuracy in the foreseeable future. There is, however, important science to be done at high resolution, work that can only be done with a large ground-based telescope at the best possible ground-based site— the Antarctic Plateau.

## 2. Development of Astronomy in the Antarctic

Water vapor is the principal source of atmospheric noise in radio observations. Because it is exceptionally cold, the climate at the South Pole implies exceptionally dry observing conditions. As air becomes colder, the amount of water vapor it can hold is dramatically reduced. At 0 C, the freezing point of water, air can hold 83 times more water vapor than saturated air at the South Pole's average annual temperature of -49 C (Goff & Gratch 1946). Together with the relatively high altitude of the Pole (2850 m), this means the water vapor content of the atmosphere above the South Pole is two or three orders of magnitude smaller than it is at most places on the Earth's surface. This has long been known (Smythe & Jackson 1977), but many years of hard work were needed to realize the potential in the form of new astronomical knowledge (cf. the recent review by Indermuehle, Burton, & Maddison 2005).

A French experiment, EMILIE  (Pajot et al. 1989), made the first astronomical observations of submillimeter-waves from the South Pole during the Austral summer of 1984-1985. EMILIE was a ground-based single-pixel bolometer dewar operating at $\lambda 900\mu m$ and fed by a 45 cm off-axis mirror. It had successfully measured the diffuse galactic emission while operating on Mauna Kea in Hawaii in 1982, but the accuracy of the result had been limited by sky noise (Pajot et al. 1986).   Martin A. Pomerantz, a cosmic ray researcher at Bartol Research Institute, encouraged the EMILIE group to relocate their experiment to the South Pole (Lynch 1998). There they found better observing conditions and were able to make improved measurements of galactic emission.

Pomerantz also enabled Mark Dragovan, then a researcher at Bell Laboratories, to attempt CMB anisotropy measurements from the Pole. Dragovan, Stark, and Pernic (1990) built a lightweight 1.2 m offset telescope and were able to get it working at the Pole with a single-pixel helium-4 bolometer during several weeks in January 1987 (See Figure 1).  The results were sufficiently encouraging that several CMB groups (Dragovan et al. 1989, Gaier et al. 1989, Meinhold et al.

1989, Peterson et al. 1989) participated in the "Cucumber" campaign in the Austral summer of 1988-1989, where three Jamesway tents and a generator were set up at a temporary site dedicated to CMB anisotropy 2 km from South Pole Station in the direction of the International Date Line. These were summer-only campaigns, where instruments were shipped in, assembled, tested, used, disassembled, and shipped out in a single three-month-long summer season. Considerable time and effort were expended in establishing and then demolishing observatory facilities, with little return in observing time. What little observing time was available occurred during the warmest and wettest days of mid-summer.

Permanent, year-round facilities were needed. The Antarctic Submillimeter Telescope and Remote Observatory (AST/RO, Stark 1997a, Stark et al. 2001) was a 1.7 m diameter offset Gregorian telescope mounted on a dedicated, permanent observatory building. It was the first radio telescope to operate year-round at South Pole. AST/RO was started in 1989 as an independent project, but in 1991 it became part of a newly-founded National Science Foundation Science and Technology Center, the Center for Astrophysical Research in Antarctica (CARA, cf. http://astro.uchicago.edu/cara and Landsberg 1998). CARA fielded several telescopes: White Dish (Tucker et al. 1993), Python (Dragovan et al. 1994, Alvarez et al 1995, Ruhl et al 1995, Platt et al. 1997, Coble et al. 1999), Viper (Peterson et al. 2000), the Degree-Angular Scale Interferometer (DASI, Leitch et al. 2002a), and the South Pole Infrared Explorer (SPIREX, Nguyen 1996), a 60-cm telescope operating primarily in the near-infrared K band. These facilities were housed in the "Dark Sector", a grouping of buildings that includes the AST/RO building, the Martin A. Pomerantz Observatory building (MAPO) and a new "Dark Sector Laboratory" (DSL), all located 1 km away from the main base across the aircraft runway in a radio quiet zone.

The combination of White Dish, Python, and UCSB 1994 (Ganga et al. 1997) data gave the first indication, by 1997, that the spectrum of spatial anisotropy in the CMB was consistent with a flat cosmology. Figure 2 shows the state of CMB anisotropy measurements as of May 1999. The early South Pole experiments, shown in green, clearly delineate a peak in CMB anisotropy at a scale $\ell = 200$, or $1^\circ$, consistent with a flat $\Omega_0 = 1$ Universe. Shortly thereafter, the BOOMERanG-98 long-duration balloon experiment (de Barnardis et al. 2000, Masi et al. 2006, 2007, Piacentini et al. 2007) and the first year of DASI (Leitch et al. 2002b), provided significantly higher signal-to-noise data, to yield $\Omega_0 = 1$ with errors less than 5%. This was a stunning achievement, definitive observations of a flat Universe balanced between open and closed Friedmann solutions. In its second year, a modified DASI made the first measurement of polarization in the CMB (Leitch et al. 2002c, Kovac et al. 2002). The observed relationship between polarization and anisotropy amplitude provided a detailed confirmation of the acoustic oscillation model of CMB anisotropy (Hu and White 1997) and strong support for the standard model. The demonstration that the geometry of the Universe is flat is an Antarctic result.

## 3. Site Testing

One of the primary tasks for the CARA collaboration was the characterization of the South Pole as an observatory site (Lane 1998). It proved unique among observatory sites for unusually low wind speeds, the complete absence of rain, and the consistent clarity of the submillimeter sky. Schwerdtfeger (1984) and Warren (1996) have comprehensively reviewed the climate of the Antarctic Plateau and the records of the South Pole meteorology office. Chamberlin (2001) analyzed weather data to determine the precipitable water vapor (PWV), a measure of total water vapor content in a vertical column through the atmosphere. He found median wintertime PWV values of 0.3 mm over a 37-year period, with little annual variation. PWV values at South Pole are small, stable, and well-understood.

Submillimeter-wave atmospheric opacity at South Pole has been measured using skydip techniques. Chamberlin, Lane, & Stark (1997) made over 1100 skydip observations at 492 GHz (λ609μm) with AST/RO during the 1995 observing season. Even though this frequency is near a strong oxygen line, the opacity was below 0.70 half of the time during the Austral winter and reached values as low as 0.34, better than ever measured at any other ground-based site. From early 1998, the λ350μm band has been continuously monitored at Mauna Kea, Chajnantor, and South Pole by identical tipper instruments developed by S. Radford of NRAO and J. Peterson of Carnegie-Mellon U. and CARA. The 350 μm opacity at the South Pole is consistently better than at Mauna Kea or Chajnantor.

Sky noise is caused by fluctuations in total power or phase of a detector caused by variations in atmospheric emissivity and path length on timescales of order one second. Sky noise causes systematic errors in the measurement of astronomical sources. This is especially important at the millimeter wavelengths, for observations of the CMB: at millimeter waves, the opacity of the atmosphere is at most a few percent, and the contribution to the receiver noise is at most a few tens of degrees, but sky noise may still set limits on observational sensitivity. Lay & Halverson (2000) show analytically how sky noise causes observational techniques to fail: fluctuations in a component of the data due to sky noise integrates down more slowly than $t^{-1/2}$ and will come to dominate the error during long observations. Sky noise at South Pole is considerably smaller than at other sites, even comparing conditions of the same opacity. The PWV at South Pole is often so low that the opacity is dominated by the dry air component (Chamberlin & Bally 1995, Chamberlin 2001); the dry air emissivity and phase error do not vary as strongly or rapidly as the emissivity and phase error due to water vapor. Lay and Halverson (2000) compared the Python experiment at South Pole (Dragovan et al. 1994, Alvarez et al. 1995, Ruhl et al. 1995, Platt et al. 1997, Coble 1999) with the Site Testing Interferometer at Chajnantor (Radford et al 1996,

Holdaway et al. 1995) and find that the amplitude of the sky noise at South Pole is 10 to 50 times less than that at Chajnantor (Bussman, Holzapfel, & Kuo 2004).

The best observing conditions occur only at high elevation angles, and at South Pole this means that only the southernmost 3 steradians of the celestial sphere are accessible with the South Pole's uniquely low sky noise— but this portion of sky includes millions of galaxies and cosmological sources, the Magellanic clouds, and most of the fourth quadrant of the Galaxy. The strength of South Pole as a millimeter and submillimeter site lies in the low sky noise levels routinely obtainable for sources around the South Celestial Pole.

## 4. Telescopes and Instruments at the South Pole

Viper was a 2.1 meter off-axis telescope designed to allow measurements of low contrast millimeter-wave sources. It was mounted on a tower at the opposite end of the MAPO from DASI. Viper was used with a variety of instruments: Dos Equis, a CMB polarization receiver operating at 7 mm, SPARO, a bolometric array polarimeter operating at λ450μm, and ACBAR, a multi-wavelength bolometer array used to map the CMB. ACBAR is a 16-element bolometer array operating at 300 mK. It was designed specifically for observations of CMB anisotropy and the Sunyaev-Zel'dovich effect (SZE). It was installed on the Viper telescope early in 2001, and was successfully operated until 2005. ACBAR has made high-quality maps of SZE in several nearby clusters of galaxies and has made significant measurements of anisotropy on the scale of degrees to arcminutes (Runyan et al. 2003, Reichardt et al. 2008).

The Submillimeter Polarimeter for Antarctic Remote Observing (SPARO) was a 9-pixel polarimetric imager operating at λ450μm. It was operational on the Viper telescope during the early Austral winter of 2000. Novak et al. (2000) mapped the polarization of a region of the sky (~0.25 square degrees) centered approximately on the Galactic Center. Their results imply that, within the Galactic Center molecular gas complex, the toroidal component of the magnetic field is dominant. The data show that all of the existing observations of large-scale magnetic fields in the Galactic Center are basically consistent with the "magnetic outflow" model of Uchida et al. (1985). This magnetodynamic model was developed in order to explain the Galactic Center Radio Lobe, a limb-brightened radio structure that extends up to one degree above the plane and may represent a gas outflow from the Galactic Center.

The Degree Angular Scale Interferometer (DASI, Leitch et al. 2002a) was a compact centimeter-wave interferometer designed to image the CMB primary anisotropy and measure its angular power spectrum and polarization at angular scales ranging from two degrees to several arcminutes. As an interferometer, DASI measured CMB power by simultaneous differencing on several scales, measuring the CMB power spectrum directly. DASI was installed on a tower adjacent to the MAPO during the 1999-2000 Austral summer and had four successful winter seasons. In its first season, DASI made measurements of CMB anisotropy that confirmed with high accuracy the "Concordance" cosmological model, which has a flat geometry and significant contributions to the total stress-energy from dark matter and dark energy (Halverson et al. 2002, Pryke 2002). In its second year, DASI made the first measurements of "E-mode" polarization of the CMB (Leitch et al. 2002c, Kovac et al. 2002).

The Antarctic Submillimeter Telescope and Remote Observatory (AST/RO) was general-purpose 1.7 m diameter telescope (Stark et al. 1997, 2001) for astronomy and aeronomy studies at wavelengths between 200 and 2000 μm. It was operational from 1995 through 2005, and was located in the Dark Sector on its own building. It was used primarily for spectroscopic studies of neutral atomic carbon and carbon monoxide in the interstellar medium of the Milky Way and the Magellanic Clouds. Six heterodyne receivers and a bolometer array were used on AST/RO: [1] a 230 GHz SIS receiver (Kooi et al. 1992), [2] a 450−495 GHz SIS quasi-optical receiver (Engargiola et al. 1994, Zmuidzinas 1992), [3] a 450−495 GHz SIS waveguide receiver (Walker et al. 1992, Kooi et al. 1995), which could be used simultaneously with [4] a 800−820 GHz fixed-tuned SIS waveguide mixer receiver (Honingh 1997), [5] the PoleSTAR array, which deployed four 800 to 820 GHz fixed-tuned SIS waveguide mixer receivers (see http://soral.as.arizona.edu/pole-star Walker et al. 2001, Groppi et al. 2000), [6] TREND, a 1.5 THz heterodyne receiver (Gerecht et al. 1999, Yngvesson et al. 2001), and [7] SPIFI, an imaging Fabry-Perot interferometer (Swain 1998). Spectral lines observed with AST/RO included: CO $J = 7 \to 6$, CO $J = 4 \to 3$, CO $J = 2 \to 1$, HDO $J = 1_{0,1} \to 0_{0,0}$, [C I] $^3P_1 \to {}^3P_0$, [C I] $^3P_2 \to {}^3P_1$, and [$^{13}$C I] $^3P_2 \to {}^3P_1$. There were four acousto-optical spectrometers (AOS, Schieder et al. 1989): two low resolution spectrometers with a bandwidth of 1 GHz, an array AOS having four low resolution spectrometer channels with a bandwidth of 1 GHz for the PoleSTAR array, and one high-resolution AOS with 60 MHz bandwidth. AST/RO produced data for over a hundred scientific papers relating to star formation in the Milky Way and the Magellanic Clouds. Among the more significant are a submillimeter-wave spectral line survey of the galactic center region (Martin et al. 2004), that showed the episodic nature of starburst and black hole activity in the center of our Galaxy (Stark et al. 2004).

QUEST at DASI (QUaD, Church et al. 2003) is a CMB polarization experiment that placed a highly-symmetric antenna feeding a bolometer array on the former DASI mount at MAPO, becoming operational in 2005. It is capable of measuring amplitude and polarization of the CMB on angular scales as small as $0.07^\circ$. QUaD has sufficient sensitivity to detect the conversion of E-mode CMB polarization to B-mode polarization caused by gravitational lensing in concentrations of Dark Matter.

BICEP (Keating et al. 2003, Yoon et al. 2006) is a millimeter-wave receiver designed to measure polarization and amplitude of the CMB over a $20^\circ$ field of view with $1^\circ$ resolution. It is mounted on the roof of the Dark Sector Laboratory, and has been operational since early 2006. The design of BICEP is optimized to eliminate systematic background effects, and thereby achieve sufficient polarization sensitivity to detect the component of CMB polarization caused by primordial gravitational waves. These measurements test the hypothesis of inflation during the first fraction of a second after the Big Bang.

The South Pole Telescope is a 10 m diameter off-axis telescope that was installed during the 2006-2007 season (Ruhl et al. 2004). It is equipped with a large field of view (Stark 2000) that feeds a state-of-the-art 936-element bolometer array receiver. The initial science goal is a large Sunyaev-Zel'dovich Effect survey covering 4000 square degrees at $1.3'$ resolution with $10\mu K$ sensitivity at a wavelength of 2 mm. This survey will find all galaxy clusters above a mass limit of $3.5 \times 10^{14} M_\odot$ regardless of redshift. It is expected that an unbiased sample of approximately 8,000 clusters will be found, with over 300 at redshifts greater than one. The sample will provide sufficient statistics to use the density of clusters to determine the equation of state of the dark energy component of the Universe as a function of time.

## 5. Conclusions

Observations from the Antarctic have brought remarkable advances in cosmology. Antarctic observations have definitively demonstrated that the geometry of the Universe is flat. These observations were made possible by excellent logistical support offered for the pursuit of science at the Antarctic bases. Cold climate and lack of water vapor provide atmospheric conditions that for some purposes is nearly as good as space, but at greatly reduced cost. Antarctica provides a platform for innovative, small instruments operated by small groups of scientists as well as telescopes that are too large to be lifted into orbit. In future, Antarctica will continue to be an important site for observational cosmology.

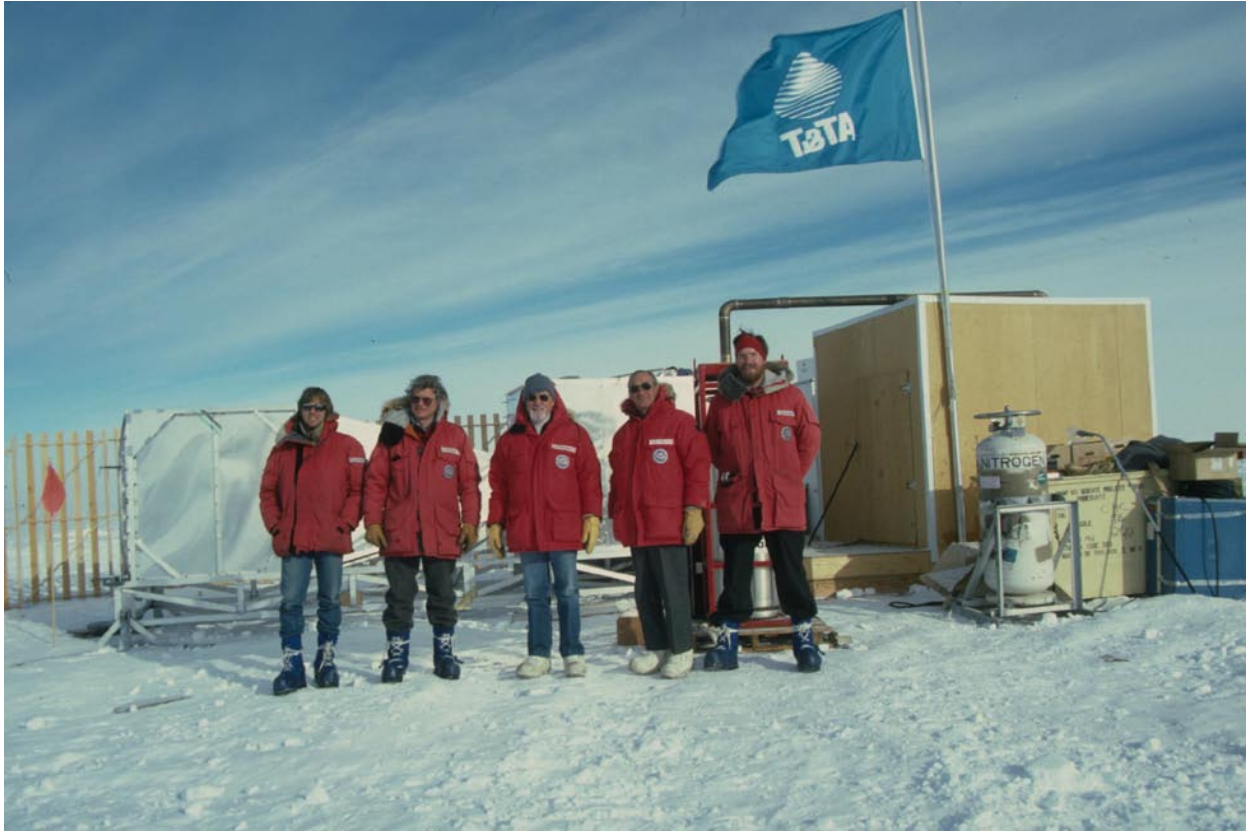

Figure 1. Mark Dragovan, Robert Pernic, Martin Pomerantz, Robert Pfeiffer, and Tony Stark, with the AT&T Bell Laboratories 1.2 meter horn antenna at the South Pole in January 1987. This was the first attempt at a CMB measurement from the South Pole.

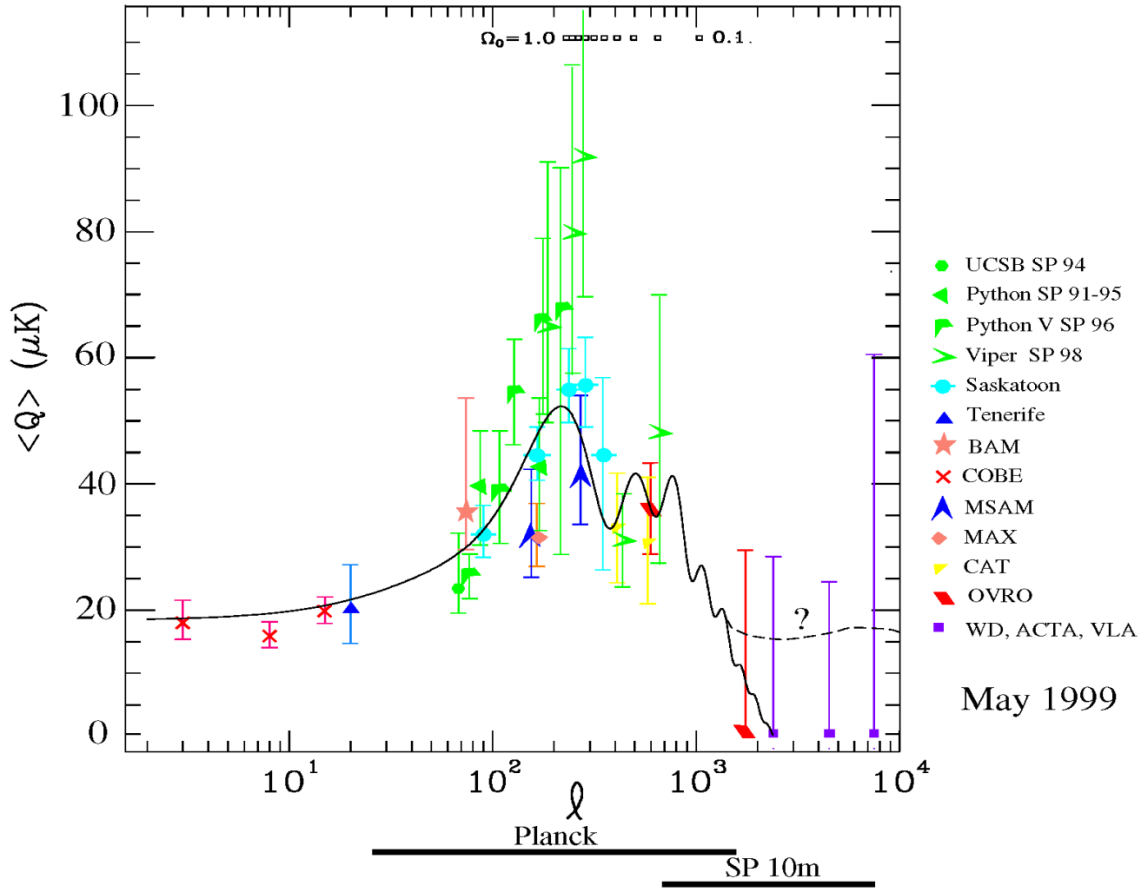

Figure 2. Microwave background anisotropy measurements as of May 1999, prior to the launch of Boomerang, the deployment of DASI , and the launch of WMAP. South Pole experimental results are shown in green. Note that the peak at $\ell = 200$ is clearly defined, indicating a flat Universe ($\Omega_0 = 1$).